\documentstyle[aps,amssymb,12pt]{revtex}
\begin{document}
\author{Yi-Hang Nie$^{1,2,}$\thanks{{\bf E-mail:} nie\_yh@yahoo.com.cn}, Yan-Hong Jin%
$^3,$J.-Q Liang$^1$,\ F.-C Pu$^{3,4}$}
\address{$^1$Institute of Theoretical Physics and Department of Physics, Shanxi\\
University,\ Taiyuan, Shanxi 030006, China\\
$^2$Department of Physics, Yanbei Normal Institute, Datong, Shanxi\\
037000,China\\
$^3$Institute of Physics and Center for Condensed Matter Physics, \\
Chinese Academy of Sciences, Beijing 100080, China\\
$^4$Department of Physics, Guangzhou Normal College, Guangzhou 510400, China}
\title{Tunnel splitting and quantum phase interference in biaxial ferrimagnetic
particles at excited states}
\maketitle

\begin{abstract}
\mathstrut \allowbreak \bigskip

The tunneling splitting in biaxial ferrimagnetic particles at excited states
with an explicit calculation of the prefactor of exponent is obtained in
terms of periodic instantons which are responsible for tunneling at excited
states and is shown as a function of magnetic field applied along an
arbitrary direction in the plane of hard and medium axes. Using complex time
path-integral we demonstrate the oscillation of tunnel splitting with
respect to the magnitude and the direction of the magnetic field due to the
quantum phase interference of two tunneling paths of opposite windings . The
oscillation is gradually smeared and in the end the tunnel splitting
monotonously increases with the magnitude of the magnetic field when the
direction of the magnetic field tends to the medium axis. The oscillation
behavior is similar to the recent experimental observation with Fe$_8$
molecular clusters. A candidate of possible experiments to observe the
effect of quantum phase interference in the ferrimagnetic particles is
proposed.
\end{abstract}

{\bf PACS number(s):} 75.30.GW, 03.65.Sq, 75.45.+j, 11.10.E5

\section{Introduction}

The macroscopic quantum phenomenon in spin system at low temperature has
attracted considerable attention both theoretically and experimentally for
more than a decade\cite{Leggett,E M Chud,Barbara,Gunther}. The magnetization
vector in a single domain ferromagnetic(FM) grain and the N\'{e}el vector in
a single domain antiferromagnetic(AFM) grain can tunnel from a metastable
state to a stable one, which is called the macroscopic quantum
tunneling(MQT), or display a coherent oscillation between two degenerate
states, which results in the superposition of macroscopically
distinguishable (classically degenerate) states ( the understanding of which
is a long-standing problem in quantum mechanics) and is called macroscopic
quantum coherence(MQC). The geometrical phase (known as the Berry phase)
interference plays a crucial role in the MQC. The quenching of MQC can be
interpreted by the quantum interference between tunneling paths of opposite
windings which possess a phase with obvious geometric meaning\cite{Loss,E M
Chud2,Liang1}. The quenching of MQC for half-integer spin has been shown
physically to be related to Kramers$^{\prime }$ degeneracy, however, the
effect of geometric phase interference is far richer than that. For example,
when the external magnetic field is applied along the hard anisotropy axis,
a new quenching of MQC occurs and is not related to Kramers$^{\prime }$
degeneracy since the external magnetic field breaks the time reversal
symmetry\cite{Garg1}. The Zeeman energy of the biaxial spin particle
associated with the external magnetic field produces an additional geometric
phase of tunnel paths which leads to the quantum interference, and the
tunnel splitting therefore oscillates with respect to the magnetic field.
The oscillations of the level splitting for the ferromagnetic particles have
been verified by the experiment with molecular clusters Fe$_8$ which at low
temperature behave like a ferromagnetic particle\cite{Wernsdorfer}. The
experimental observation of the oscillation of tunnel splitting has
triggered off more detailed investigations along this direction\cite
{Garg2,KouSP,JinYH}. Since the tunneling rate in AFM particles is much
higher than that in FM particles of the same volume\cite{Duan} the AFM
particles are expected to be a better candidate for the observation of
macroscopic quantum phenomena than the FM particles. The quantum tunneling
of the N\'{e}el vector in AFM particles has been well studied in terms of
the idealized sublattice-model\cite{Garg1,Simanujunlak} in which the
external magnetic field does not play a role since the net magnetic moment
vanishes. The biaxial AFM particles with a small noncompensation of
sublattices or in other words biaxial ferrimagnetic particles have to be
considered in order to obtain the effect of the external magnetic field on
the tunnel splitting. The oscillation of tunnel splitting at ground state of
the biaxial ferrimagnetic particles was predicted recently with the magnetic
field applied along the hard axis \cite{Nie1}. In the present paper we
investigate the effect of quantum phase interference at excited states for a
biaxial ferrimagnetic particle in the external magnetic field applied along
an arbitrary direction in the plane of hard and medium axis. Since the
effect of geometric phase interference has been observed in the experiment
of Fe$_8$ molecular clusters with the magnetic field along an arbitrary
direction, the present generalization to the ferrimagnetic particles is not
only of theoretical but also of practical interests. At ground state one
only considers the paths of imaginary time under barrier. The extension to
excited states is highly nontrivial. Paths of complex time have to be taken
into account since a path at excited states also approaches the region of
potential well and therefore is of real time.

\section{ effective Lagrangian of a biaxial ferrimagnetic particle in a
magnetic field}

We consider a biaxial AFM particle of two collinear FM sublattices with a
small non-compensation. Assuming that the particle possesses a X easy axis
and XOY easy plane , and the magnetic field $h$ is applied along an
arbitrary direction in the plane of the hard axis (Z axis) and medium axis(Y
axis), the Hamiltonian operator of the AFM particle has the form

\begin{equation}
\hat{H}=\sum_{a=1,2}\left( k_{\bot }\hat{S}_{{\small a}}^{z2}+k_{%
\shortparallel }\hat{S}_a^{y2}-\gamma h_z\hat{S}_a^z-\gamma h_y\hat{S}%
_a^y\right) +J\hat{S}_1\cdot \hat{S}_2,
\end{equation}
where $k_{\bot },k_{\shortparallel }>0$ are the anisotropy constants, $J$ is
the exchange constant, $\gamma $ is the gyromagnetic ratio, and the spin
operators in two sublattices $\hat{S}_1$ and $\hat{S}_2$ obey the usual
commutation relation $\left[ \hat{S}_a^i,\hat{S}_b^j\right] =i\hbar \epsilon
_{ijk}\delta _{ab}\hat{S}_b^k\ \left( i,j,k=x,y,z;a,b=1,2\right) $. In order
to obtain the Lagrangian of the system, we begin with the matrix element of
the evolution operator in spin coherent-state representation by means of the
spin coherent state path integrals 
\begin{equation}
\langle N_f|e^{-2i\hat{H}T/\hbar }|\text{$N$}_i\rangle =\int \left[
\prod_{k=1}^{M-1}d\mu \left( \text{$N$}_k\right) \right] \left[
\prod_{k=1}^M\langle \text{$N$}_k|e^{-i\epsilon \hat{H}/\hbar }|\text{$N$}%
_{k-1}\rangle \right] .
\end{equation}

Here we define $|N\rangle =|n_1\rangle |n_2\rangle $, $|N_M\rangle
=|N_f\rangle =|n_{1,f}\rangle |n_{2,f}\rangle ,$ $|N_0\rangle =|N_i\rangle
=|n_{1,i}\rangle |n_{2,i}\rangle $, $t_f-t_i=2T$ and $\epsilon =2T/M$. The
spin coherent state is defined as 
\begin{equation}
|n_a\rangle =e^{i\theta _a\hat{O}_a}|S_a,S_a\rangle ,\left( a=1,2\right) 
\end{equation}
where $n_a=\left( \sin \theta _a\text{cos}\phi _a,\text{sin}\theta _a\text{%
sin}\phi _a,\text{cos}\theta _a\right) $ is the unit vector, $\hat{O}_a=$sin$%
\phi _a\hat{S}_a^x-$cos$\phi _a\hat{S}_a^y$ and $|S_{a\text{,}}S_a\rangle $
is the reference spin eigenstate. The measure is defined by 
\begin{equation}
d\mu \left( \text{$N$}_k\right) =\prod_{a=1,2}\frac{2S_a+1}{4\pi }\text{sin}%
\theta _{a,k}d\theta _{a,k}d\phi _{a,k},
\end{equation}
Evaluating the path integral on the right hand side of the Eq.(2) we obtain
in the large $S$ limit\cite{Nie2} 
\begin{equation}
\langle \text{$N$}_f|e^{-2i\hat{H}T/\hbar }|\text{$N$}_i\rangle =\int
\prod_{a=1,2}D[\theta _a]D[\phi _a]\exp \left[ \frac i\hbar
\int_{t_i}^{t_f}\left( L_0+L_1\right) dt\right] 
\end{equation}
with 
\begin{equation}
L_0=\sum_{a=1,2}S_a\dot{\phi}_a(\cos \theta _a-1)~-JS_1S_2\left[ \sin \theta
_1\sin \theta _2\cos \left( \phi _1-\phi _2\right) +\cos \theta _1\cos
\theta _2\right] ,
\end{equation}
\begin{equation}
L_1=-\sum_{a=1,2}\left( k_{\bot }S_a^2\cos ^2\theta _a+k_{\shortparallel
}S_a^2\sin ^2\theta _a\sin ^2\phi _a-\gamma h_zS_a\cos \theta _a-\gamma
h_yS_a\sin \theta _a\sin \phi _a\right) ,
\end{equation}
where $L_0+L_1$ denotes the Lagrangian. Sin\ ce spins $S_1$ and $S_2$ in two
sublattices are almost antiparallel, we may replace $\theta _2$ and $\phi _2$
by $\theta _2=\pi -\theta _1-\epsilon _{_\theta }$and $\phi _2=\pi +\phi
_1+\epsilon _{_\phi },$ where $\epsilon _{_\theta }$ and $\epsilon _{_\phi }$
denote small fluctuations. Working out the fluctuation integrations over $%
\epsilon _{_\theta }$ and $\epsilon _{_\phi }$ the transition amplitude
Eq.(5) reduces to 
\begin{equation}
\langle \text{$N$}_f|e^{-2i\hat{H}T/\hbar }|\text{$N$}_i\rangle =\int
D[\theta ]D[\phi ]\exp \left( \frac i\hbar \int_{t_i}^{t_f}\bar{L}dt\right) ,
\end{equation}
\begin{equation}
\bar{L}=\Omega \left[ -\frac{M_1+M_2}\gamma \dot{\phi}+\frac M\gamma \dot{%
\phi}\cos \theta +\frac{\chi _{\bot }}{2\gamma ^2}\left( \dot{\theta}^2+\dot{%
\phi}^2\sin ^2\theta \right) \right] -V\left( \theta ,\phi \right) ,
\end{equation}
where $V\left( \theta ,\phi \right) =\Omega K_{\bot }\left( \cos \theta
-Mh_z/2K_{\bot }\right) ^2+\Omega K_{\shortparallel }\sin ^2\theta \left(
\sin \phi -Mh_y/2K_{\shortparallel }\sin \theta \right) ^2,$ and $\left(
\theta _1,\phi _1\right) $ has been replaced by $\left( \theta ,\phi \right) 
$. $M_a=\gamma \hbar S_a/\Omega \ (a=1,2),\ M=\gamma \hbar \left(
S_1-S_2\right) /\Omega $ with $\Omega $ being the volume of the AFM particle
and $\chi _{\bot }=\gamma ^2/J$. $K_{\bot }=2k_{\bot }S^2/\Omega $ and $%
K_{\shortparallel }=2k_{\shortparallel }S^2/\Omega $ (setting $S_1=S_2=S$
except in the term containing $S_1-S_2$) denote the transverse and the
longitudinal anisotropy constants, respectively.

We assume a very strong transverse anisotropy, $i.e.,$ $K_{\bot }\gg $ $%
K_{\shortparallel }$ . For this case, the N\'{e}el vector is forced to lie
on a cone of angle $2\theta _0$. Where $\cos \theta _0=Mh_z/2K_{\bot
}=\delta h_z/h_c\ (\delta =K_{\shortparallel }/K_{\bot },\
h_c=2K_{\shortparallel }/M$ ). Introducing the fluctuation variable $\eta $
such that $\theta =\theta _0+$ $\eta $ and considering $K_{\bot }\gg $ $%
K_{\shortparallel }$we have $V(\theta ,\phi )=\Omega K_{\bot }\sin ^2\theta
_0\eta ^2+\Omega K_{\shortparallel }\sin ^2\theta _0(\sin \phi -b)^2\
(b=Mh_y/2K_{\shortparallel }\sin \theta _0=h\sin \alpha /\delta \sqrt{%
(h_c/\delta )^2-h_z^2}$ $(\sin ^2\theta _0=1-(\delta h_z/h_c)^2),$ where $%
\alpha $ is the angle between the magnetic field and Z axis$)$ and thus the
Eq.(9) is written as

\begin{eqnarray}
\bar{L} &=&\Omega \left[ \frac 12\left( \frac{M^2}{2K_{\bot }\gamma ^2}+%
\frac{\chi _{\bot }}{\gamma ^2}\right) \dot{\phi}^2-\frac{M_1+M_2}\gamma 
\dot{\phi}+\frac M\gamma \dot{\phi}\cos \theta _0-K_{\shortparallel }\sin
^2\theta _0(\sin \phi -b)^2\right]   \nonumber \\
&&+\Omega \left[ \frac{\chi _{\bot }}{2\gamma ^2}\ \dot{\eta}^2-K_{\bot
}\sin ^2\theta _0\left( \eta +\frac{M\dot{\phi}}{2K_{\bot }\gamma \sin
\theta _0}\right) ^2\right] .
\end{eqnarray}
Carrying out the integral over $\eta $ we obtain

\begin{equation}
\langle \text{$N$}_f|e^{-2\hat{H}\beta /\hbar }|\text{$N$}_i\rangle =\int
D[\phi ]\exp \left( -\frac 1\hbar \int_{\tau _i}^{\tau _f}L_{eff}d\tau
\right)
\end{equation}
where

\begin{equation}
L_{eff}=\frac I2\left( \frac{d\phi }{d\tau }\right) ^2+i\Theta \frac{d\phi }{%
d\tau }+V\left( \phi \right) 
\end{equation}
is the effective Euclidean Lagrangian. $\tau =it$ and $\beta =iT$ . $%
I=I_a+I_f$ where $I_f=\Omega M^2/(2\gamma ^2K_{\bot })$ and $I_a=\Omega \chi
_{\bot }\sin ^2\theta _0/\gamma ^2$ are the effective FM and AFM moments of
inertia\cite{E.M.Chudnovsky 3M}, respectively. $V\left( \phi \right) =\Omega
K_{\shortparallel }\sin ^2\theta _0(\sin \phi -b)^2$ is the effective
potential and $\Theta =\hbar (S_0-d)$ $(S_0=S_1+S_2$ and $d=h_z/h_0=h\cos
\alpha /h_0$ with $h_0=\hbar /\gamma I_f)$. The second term in the Eq.(12) , 
$i.e.$, $i\Theta \frac{d\phi }{d\tau }$ has no effect on the classical
equation of motion, however, it leads to a path dependent phase in Euclidean
action. When $h_y$=0, $V(\phi )=K_{\shortparallel }\Omega \sin ^2\theta
_0\sin ^2\phi $ possesses the form of the sin-Gordon potential and the
directions with $\theta =\theta _0$, $\phi =0$ and $\pi $ are two
equilibrium orientations of the N\'{e}el vector(Fig.1(b)) around which the
small oscillation frequency of the N\'{e}el vector is seen to be $\omega _0=%
\sqrt{2K_{\shortparallel }\Omega \sin ^2\theta _0/I}.$ The quantum tunneling
of the N\'{e}el vector through two paths of opposite windings results in the
quantum phase interference. When $h_y\neq 0,$ the potential $V(\phi )=\Omega
K_{\shortparallel }\sin ^2\theta _0(\sin \phi -b)^2$ has an asymmetric
twin-barrier (Fig.2(a)), and the net magnetic moment of the uncompensated
sublattices in the applied magnetic field shifts the equilibrium
orientations of the N\'{e}el vector to $\phi =\phi _{+}$ and $\pi -\phi _{+%
\text{ }}\ (\phi _{+}=\arcsin b)$(Fig.2(b)) around which the small
oscillation frequency of the N\'{e}el vector is modified as $\omega =\omega
_0\sqrt{1-b^2}.$ The quantum tunneling of the N\'{e}el vector through two
different barriers leads to the quantum phase interference.

\section{Quantum phase interference as h$_y$=0}

When the external magnetic field is applied along the hard axis(Z axis), the
effective potential is $V(\phi )=\Omega K_{\shortparallel }\sin ^2\theta
_0\sin ^2\phi $ . The quantum tunneling at finite energy $E$ is dominated by
the periodic instantons\cite{Manton}. From the Euclidean Lagrangian (12),
the equation of motion of the pseudoparticles moving in the classically
forbidden region in the barrier  is seen to be

\begin{equation}
\frac I2\left( \frac{d\phi }{d\tau }\right) ^2-V(\phi )=-E.
\end{equation}
The N\'{e}el vector may rotate by tunneling through potential barriers from
one orientation($\phi =0$) to another($\phi =\pi $) along clockwise path and
anticlockwise path (Fig.1). The instantons satisfying periodic boundary
condition are found to be

\begin{equation}
\phi _c^{\pm }=\pm \frac \pi 2\pm \arcsin \left[ k_1\text{sn(}\omega _0\tau 
\text{)}\right] 
\end{equation}
where ''-'' denotes the clockwise path and ''+'' denotes the anticlockwise
path(see Fig.1), sn($\omega _0\tau $) is the Jacobian elliptic function with
modulus

\begin{equation}
k_1=\sqrt{1-\frac E{\Omega K_{\shortparallel }\sin ^2\theta _0}}.
\end{equation}
The two trajectories of instantons $\phi _c^{\pm }$ are shown in Fig.1(a).
The Euclidean actions evaluated along the trajectories of periodic
instantons are 
\begin{equation}
S_e^{\pm }=W_e+2E\beta +i\theta _e^{\pm },
\end{equation}

\begin{equation}
W_e=\int_{-\beta }^\beta \left[ \frac I2\left( \frac{d\phi _c^{\pm }}{d\tau }%
\right) ^2-V(\phi _c^{\pm })\right] d\tau =\frac{4\Omega K_{\shortparallel
}\sin ^2\theta _0}{\omega _0}\left[ E(k_1)-k_1^{^{\prime }2}{\cal K}%
(k_1)\right] ,
\end{equation}
\begin{equation}
\theta _e^{\pm }=\int_{-\beta }^\beta \Theta \frac{d\phi _c^{\pm }}{d\tau }%
d\tau =\pm \Theta (\pi -2\arcsin k_1^{^{\prime }})
\end{equation}
where ${\cal K}(k_1),\ E(k_1)$ are the complete elliptic integrals of the
first and the second kinds, respectively. $k_1^{^{\prime
}2}=1-k_1^2=E/\Omega K_{\shortparallel }\sin ^2\theta _0$. To investigate
the quantum tunneling and related quantum phase interference at excited
states, we begin with the instanton induced transition amplitude

\begin{equation}
\sum_{m,n}\langle E_n^f|\hat{P}_E|E_m^i\rangle =\int d\phi _fd\phi _i\psi
_E^{*}(\phi _f)\psi _E(\phi _i)G(\phi _f,\beta ;\phi _i,-\beta ).
\end{equation}
$\hat{P}_E$ is the operator of projection onto the subspace of fixed energy 
\cite{Kuznetsov}. $|E_E^f\rangle $ and $|E_E^i\rangle $ are two excited
states lying on different sides of the barrier. From Eq.(19) the tunnel
splitting is written as

\begin{equation}
{\em \Delta E\sim }\frac{\exp \left( \frac{2E\beta }\hbar \right) }\beta
\left| \int d\phi _fd\phi _i\psi _E^{*}(\phi _f)\psi _E(\phi _i)G(\phi
_f,\beta ;\phi _i,-\beta )\right| ,
\end{equation}
\begin{equation}
G=\int {\cal D[\phi ]}\exp \left( -\frac 1\hbar \int_{-\beta }^\beta
L_{eff}d\tau \right) .
\end{equation}
When the quantum phase interference of tunneling through clockwise and
anticlockwise paths is taken into account, the Eq.(20) is written as

\begin{equation}
{\em \Delta E\sim }\frac{\exp \left( \frac{2E\beta }\hbar \right) }\beta
\left| I_1^{+}+I_1^{-}\right| ,
\end{equation}

\begin{equation}
I_1^{\pm }=\int d\phi _f^{\pm }d\phi _i^{\pm }\psi _E^{*}(\phi _f^{\pm
})\psi _E(\phi _i^{\pm })G(\phi _f^{\pm },\beta ;\phi _i^{\pm },-\beta
)=\exp \left( -\frac{i\theta _e^{\pm }}\hbar \right) I_0,
\end{equation}

\begin{equation}
I_0=\int d\phi _fd\phi _i\psi _E^{*}(\phi _f)\psi _E(\phi _i)\bar{G}(\phi
_f,\beta ;\phi _i,-\beta ),
\end{equation}

\begin{equation}
\bar{G}=\int {\cal D[\phi ]}\exp \left( -\frac 1\hbar \int_{-\beta }^\beta 
\bar{L}_{eff}d\tau \right) ,
\end{equation}
\begin{equation}
\bar{L}_{eff}=\frac I2\left( \frac{d\phi }{d\tau }\right) ^2+V(\phi ).
\end{equation}
$I_0$ is independent of tunnel directions. The phase independent tunneling
kernel $\bar{G}$ is now evaluated with the help of the periodic instantons.
Following the procedure of the periodic instanton-calculation in Refs.(20)
and (21) a general formula for Eq.(24) is found to be

\begin{equation}
I_0{\em \sim }2\beta \exp \left( -\frac{2E\beta }\hbar \right) \left[ \frac{%
\hbar \omega _0}{4{\cal K}(k_1^{^{\prime }})}\right] \exp \left( -\frac{W_e}%
\hbar \right) .
\end{equation}
To investigate the quantum phase interference at excited state, we have to
consider additional phases coming from the real-time paths in the potential
well between $0\rightarrow \phi _{1\text{ }}$and $\phi _2\rightarrow \pi $ ($%
0\rightarrow -\phi _{1\text{ }}$and $-\phi _2\rightarrow -\pi $). Thus
tunnel splitting Eq.(22) is rewritten as

\begin{eqnarray}
{\em \Delta E} &=&\frac{\exp \left( \frac{2E\beta }\hbar \right) }\beta
\left| I_1^{+}\exp \left( \frac{iS_r^{+}}\hbar \right) +I_1^{-}\exp \left( 
\frac{iS_r^{-}}\hbar \right) \right|  \nonumber \\
&=&\frac{\exp \left( \frac{2E\beta }\hbar \right) }\beta I_0\left| \exp
\left[ \frac{i(S_r^{+}-\theta _e^{+})}\hbar \right] +\exp \left[ \frac{%
i(S_r^{-}-\theta _e^{-})}\hbar \right] \right| ,
\end{eqnarray}
where

\begin{equation}
S_r^{\pm }=\theta _r^{\pm }+W_r^{\pm },
\end{equation}
\begin{equation}
\theta _r^{\pm }=-\Theta \int_{\left[ 0,\pm \phi _1\right] \cup \left[ \pm
\phi _2,\pm \pi \right] }d\phi =\mp 2\Theta \arcsin k_1^{^{\prime }},
\end{equation}
\begin{equation}
W_r^{\pm }=\pm \sqrt{\frac I2}\int_{\left[ 0,\pm \phi _1\right] \cup \left[
\pm \phi _2,\pm \pi \right] }\frac{E-2V(\phi )}{\sqrt{E-V(\phi )}}d\phi .
\end{equation}
It is obvious that $W_r^{+}=W_r^{-}$. Substituting Eqs.(27), (29) and (30)
into the Eq.(28), we obtain the tunnel splitting

\begin{equation}
{\em \Delta E=}\frac{\omega _0\hbar }{{\cal K(}k_1^{^{\prime }}{\cal )}}\exp
\left( -\frac{W_e}\hbar \right) \left| \cos (\Lambda \pi )\right|
\end{equation}
where $\Lambda =S_0-d.$ The tunnel splitting ${\em \Delta E}$ is a function
of the external magnetic field and energy.

For low lying excited states($k_1^{^{\prime }}=\sqrt{E/\Omega
K_{\shortparallel }\sin ^2\theta _0}<<1$) in which we are interested, the
energy $E$ may be replaced by the harmonic oscillator approximated
eigenvalues $E_m=(m+\frac 12)\omega _0\hbar $. Expanding the complete
elliptic integrals ${\cal K(}k_1{\cal )}$ and $E(k_1)$ as power series of $%
k^{^{\prime }}$ and taking note of limit ${\cal K(}k_1^{^{\prime
}}\rightarrow 0{\cal )\rightarrow }\frac \pi 2,$ we obtain the tunnel
splitting of the $m$th excited state,

\begin{equation}
{\em \Delta E}_m{\em =}\frac{(4B)^m}{m!}\Delta E_0|\cos (\Lambda \pi )|,
\end{equation}
where

\begin{equation}
{\em \Delta E}_0=\frac{2\hbar \omega _0}{\sqrt{\pi }}(8B)^{\frac 12}\exp (-B)
\end{equation}
with $B=4K_{\shortparallel }\Omega \sin ^2\theta _0/\hbar \omega _0$ which
denotes the tunnel splitting of ground state. It may be worth to estimate
the range of validity of our results, $i.e.$, how large $m$ is. $\Omega
K_{\shortparallel }\sin ^2\theta _0$ is the barrier height of potential and $%
\hbar \omega _0$ is the level space between neighboring levels. For the
horse-spleen ferritin reported in \cite{D.D.A.,J.Tejada} the residual spin
is $S\sim 100$ ( corresponding moment $M_0=217\mu _B$) and volume is $\Omega
\sim 2\times 10^{-19}cm^3$ (diameter 7.5nm). The longitudinal anisotropy
constant and transverse susceptibility are seen to be $K_{\shortparallel
}=2\times 10^5erg/cm^3$ and $\chi _{\bot }=10^{-5}emu/G$ $cm^3$
respectively. Using the above parameters we find that the number of the
levels in the potential well is about 10 as $\delta \sim 0.03$. Fig.3(a)
shows the oscillation of tunnel splittings of lowest 3 states with respect
to the external magnetic field due to the quantum phase interference of two
tunneling paths of opposite windings for $S_0=$integer and half-integer.
From Fig.3(a) one can find that the magnitude of tunnel splittings at
excited states is much higher than that at ground state and may contribute
significantly to the experimental observation at finite temperature. When $%
d=S_0-l-\frac 12,\ i.e.,\ h=(S_0-l-\frac 12)h_0$ ($l$ is an integer), the
tunneling splitting ${\em \Delta E}_m$ vanishes. The period of oscillation is

\begin{equation}
{\em \Delta h}=\frac \hbar {\gamma I_f}
\end{equation}
which is independent of the energy.

\section{ Quantum phase interference as h$_y\neq 0$}

When the external magnetic field is applied along an arbitrary direction in
the plane of the hard axis and medium axis, the effective potential $V(\phi
)=\Omega K_{\shortparallel }\sin ^2\theta _0(\sin \phi -b)^2$ has the
asymmetric twin barriers which lead to that N\'{e}el vector may rotate from
one orientation($\phi =\phi _{+}$) to another ($\phi =\pi -\phi _{+}$) along
clockwise underbarrier path and anticlockwise path(Fig.2). Two different
instantons (Fig.2) corresponding to tunneling through two types of barriers
are found as

\begin{equation}
\phi _c^{\pm }=\pm \frac \pi 2\pm 2\arctan \left[ \lambda _{\pm }\text{sn}%
(q\tau ,k_2)\right]
\end{equation}
where

\[
k_2=\left[ \frac{(1-\varepsilon )^2-b^2}{(1+\varepsilon )^2-b^2}\right] ^{%
\frac 12},\varepsilon =\sqrt{\frac E{\Omega K_{\shortparallel }\sin ^2\theta
_0}}, 
\]

\[
q=\frac{\omega _0}2\left[ (1+\varepsilon )^2-b^2\right] ^{\frac 12},\lambda
_{\pm }=\left[ \frac{(1-\varepsilon )^2-b^2}{(1\pm b)^2-\varepsilon ^2}%
\right] ^{\frac 12}. 
\]
Our starting point for investigation of the tunneling and related quantum
phase interference at excited states is still the transition amplitude of
the barrier penetration projected onto the subspace of fixed energy $E$, $%
i.e.,$ the Eq.(19) from which the tunneling splitting is obtained as

\begin{equation}
{\em \Delta E\sim }\frac{\exp \left( \frac{2E\beta }\hbar \right) }\beta
\left| \int d\phi _fd\phi _i\psi _E^{*}(\phi _f)\psi _E(\phi _i)G(\phi
_f,\beta ;\phi _i,-\beta )\right| ,
\end{equation}
The result corresponding to the Eq.(22) is formally the same

\begin{equation}
{\em \Delta E\sim }\frac{\exp \left( \frac{2E\beta }\hbar \right) }\beta
\left| I_2^{+}+I_2^{+}\right| .
\end{equation}
In the present case , however, the two tunneling paths are not symmetric.
Thus we find

\begin{equation}
I_2^{\pm }=\exp \left( -\frac{i\delta _e^{\pm }}\hbar \right) \bar{I}_2^{\pm
},
\end{equation}

\begin{equation}
\delta _e^{\pm }=\pm \Theta \left[ \pi \mp 2\arcsin (b\pm \varepsilon
)\right] ,
\end{equation}

\begin{equation}
\bar{I}_2^{\pm }=\int d\phi _f^{\pm }d\phi _i^{\pm }\psi _E^{*}(\phi _f^{\pm
})\psi _E(\phi _i^{\pm })\bar{G}^{\pm }(\phi _f^{\pm },\beta ;\phi _i^{\pm
},-\beta ),
\end{equation}
\begin{equation}
\bar{G}^{\pm }=\int {\cal D[\phi ]}\exp \left( -\frac 1\hbar \int_{-\beta
}^\beta \bar{L}_{eff}^{\pm }d\tau \right) ,
\end{equation}
\begin{equation}
\bar{L}_{eff}=\frac I2\left( \frac{d\phi _c^{\pm }}{d\tau }\right) ^2+V(\phi
_c^{\pm }).
\end{equation}
$\bar{I}_2^{\pm }$ is now dependent on tunnel direction. The phase dependent
tunneling kernel $\bar{G}^{\pm }$ is evaluated with the help of the periodic
instanton. Following the procedure above we obtain

\begin{equation}
\bar{I}_2^{\pm }{\em \sim }2\beta \exp \left( -\frac{2E\beta }\hbar \right)
\left[ \frac{\hbar \omega _0}{4\sigma {\cal K}(k_2^{^{\prime }})}\right]
\exp \left( -\frac{W_e^{\pm }}\hbar \right) ,
\end{equation}
\[
\sigma =\left[ (1+\varepsilon )^2-b^2\right] ^{-\frac 12},\ \ k_2^{^{\prime
}}=\sqrt{1-k_2},
\]

\begin{equation}
W_e^{\pm }=\frac{4Iq}{\lambda _{\pm }^2}\left[ \lambda _{\pm
}^2E(k_2)+(k_2^2-\lambda _{\pm }^2){\cal K}(k_2)-(\lambda _{\pm
}^4-k_2^2)\Pi (k_2,\lambda _{\pm }^2)\right] ,
\end{equation}
where $\Pi (k_2,\lambda _{\pm }^2)$ is the complete elliptic integral of the
third kind. Considering the additional phase contribution from the real-time
paths in potential well the tunnel splitting Eq. (38) is written as

\begin{equation}
{\em \Delta E}=\frac{\exp \left( \frac{2E\beta }\hbar \right) }\beta \left|
I_2^{+}\exp \left( \frac{iS_r^{+}}\hbar \right) +I_2^{-}\exp \left( \frac{%
iS_r^{-}}\hbar \right) \right| ,
\end{equation}
where

\begin{equation}
S_r^{\pm }=\delta _r^{\pm }+\Phi _r^{\pm },
\end{equation}
\begin{equation}
\delta _r^{\pm }=-2\Theta \left[ \arcsin (b\pm \varepsilon )-\arcsin
b\right] ,
\end{equation}
\begin{equation}
\Phi _r^{\pm }=2I\omega _0\left[ \frac{E(\varphi ^{\pm },k_2^{^{\prime }})}%
\sigma +\sigma (1\mp b)^2F(\varphi ^{\pm },k_2^{^{\prime }})\mp 2b\sigma
(1\mp b-\varepsilon )\Pi (\varphi ^{\pm },\alpha ^{\pm },k_2^{^{\prime
}})-\varepsilon \sqrt{\frac{1\pm b}{1\mp b}}\right] ,
\end{equation}
\[
\varphi ^{\pm }=\arcsin \sqrt{\frac{1\mp b+\varepsilon }{2(1\mp b)}},\alpha
^{\pm }=\sqrt{\frac{2\varepsilon }{1\mp b+\varepsilon }}. 
\]
Inserting Eqs.(39), (44) and (47) into Eq.(46), we obtain the final formula
of the tunnel splitting

\begin{eqnarray}
{\em \Delta E} &=&\frac{\hbar \omega _0}{2\sigma {\cal K}(k_2^{^{\prime }})}
\nonumber \\
&&\left\{ \exp \left( -\frac{2W_e^{+}}\hbar \right) +\exp \left( -\frac{%
2W_e^{-}}\hbar \right) +2\exp \left( -\frac{W_e^{+}+W_e^{-}}\hbar \right)
\cos [2\Lambda \pi -(\Phi _r^{+}-\Phi _r^{-})]\right\} ^{\frac 12}
\end{eqnarray}
which is a function of the external magnetic field and the energy. For low
lying excited states, $\varepsilon <<1$ , $k_2^{^{\prime }}<<1$, the energy $%
E$ is again replaced by harmonic oscillator approximated eigenvalues $E_m=(m+%
\frac 12)\hbar \omega $. Expanding the complete elliptic integrals $E(k_2)$, 
${\cal K(}k_2{\cal )}$ and $\Pi (k_2,\lambda _{\pm }^2)$ in the Eq.(45) as
power series of $k_2^{^{\prime }}$ we obtain

\begin{equation}
W_e^{\pm }=\frac{4\Omega K_{\shortparallel }\sin ^2\theta _0}{\omega _0}%
\left[ \sqrt{1-b^2}-\frac 1{16}\left( 1-b^2\right) ^{\frac 32}k_2^{^{\prime
}4}\left( \ln \frac 4{k_2^{^{\prime }}}+\frac 14\right) +b\arcsin b\mp \frac %
\pi 2\right] .
\end{equation}
Substituting the Eq.(51) into the Eq.(50) and taking note of limits ${\cal K(%
}k_2^{^{\prime }}\rightarrow 0{\cal )\rightarrow }\frac \pi 2$, $\sigma
\left( k_2^{^{\prime }}\rightarrow 0\right) \rightarrow (1-b^2)^{\frac 12}$
and  $\Phi _r^{+}\approx \Phi _r^{-}$ at low lying excited states we obtain
the tunnel splitting of the $m$th excited state as

\begin{equation}
{\em \Delta E}_m=\frac{E_2}{m!}\left[ 4B\left( 1-b^2\right) ^{\frac 32%
}\right] ^m\left[ \cosh (bB\pi )+\cos (2\Lambda \pi )\right] ^{\frac 12},
\end{equation}
\begin{equation}
E_2=\frac{2\hbar \omega _0}{\sqrt{\pi }}\left[ 4B\left( 1-b^2\right) ^{\frac %
52}\right] ^{\frac 12}\exp \left[ -B\left( \sqrt{1-b^2}+b\arcsin b\right)
\right] .
\end{equation}
Fig.3 shows the oscillation of tunnel splitting at low lying excited states
with respect to the external magnetic field for $S_0=$integer and
half-integer respectively. When $\Lambda =(2l+1)/2\ (l$ is an integer$),$
tunnel splitting ${\em \Delta E}_m$ tends to a minimum value. The period of
oscillation is

\begin{equation}
{\em \Delta h=}\frac{h_0}{\cos \alpha }
\end{equation}
which is independent of the level, but dependent on the direction of the
external magnetic field. When $\alpha =0$ and $m=0$, the tunnel splitting $%
{\em \Delta E}_m$ reduces to the result in Ref.[15]. The period increases
with the angle $\alpha $. When the direction of the magnetic field is along
the medium axis ($\alpha =\frac \pi 2$), the period approaches to infinity,
in other words, the oscillation disappears.

\section{conclusion}

The effect of the macroscopic quantum phase interference at excited states
is studied for the biaxial ferrimagnetic particles with the external
magnetic field applied along an arbitrary direction in the plane of hard and
medium axis. We present a general formula of tunnel splitting at excited
states as a function of the magnetic field and the energy. The oscillation
behavior of tunneling splitting at low lying excited states is similar to
that in FM particles observed experimentally in molecular clusters Fe$_8$
and should be observed in further experiment with ferrimagnetic particles
for which a possible candidate of materials may be horse-spleen ferritin\cite
{D.D.A.,J.Tejada}.

\begin{description}
\item  \bigskip \bigskip 
\end{description}

\smallskip

{\bf Acknowledgments}

This work was supported by the National Nature Science Foundation of China
under Grant No.19775033, 10075032 and Shanxi Nature Science Foundation.
Yihang Nie also acknowledges support of Yanbei Normal Institute.

\bigskip {\bf Figure caption}

Fig.1 (a) The periodic potential and the instanton trajectories. The arrow
lines denote two tunnel paths of opposite windings. (b) The equilibrium
orientations of N\'{e}el vector in the absence of Y-component of the
magnetic field.

Fig.2 (a) The potential with asymmetric twin-barrier and instanton
trajectories. (b) The equilibrium orientations of N\'{e}el vector in the
presence of Y-component of the magnetic field.

Fig.3 The level splitting as function of the external magnetic field with
angular (a) $\alpha =0^{\circ },$ (b) $\alpha =3^{\circ },$(c)$\alpha
=5^{\circ }$ for $S_0=$integer(solid line) and $S_0=$half-integer(dot line).
Here $S=100,\Omega =10^{-19}cm^3,\chi _{\bot }=10^{-5}$ , $K_{\shortparallel
}=10^5erg/cm^3$ and $\delta \sim 0.03$.

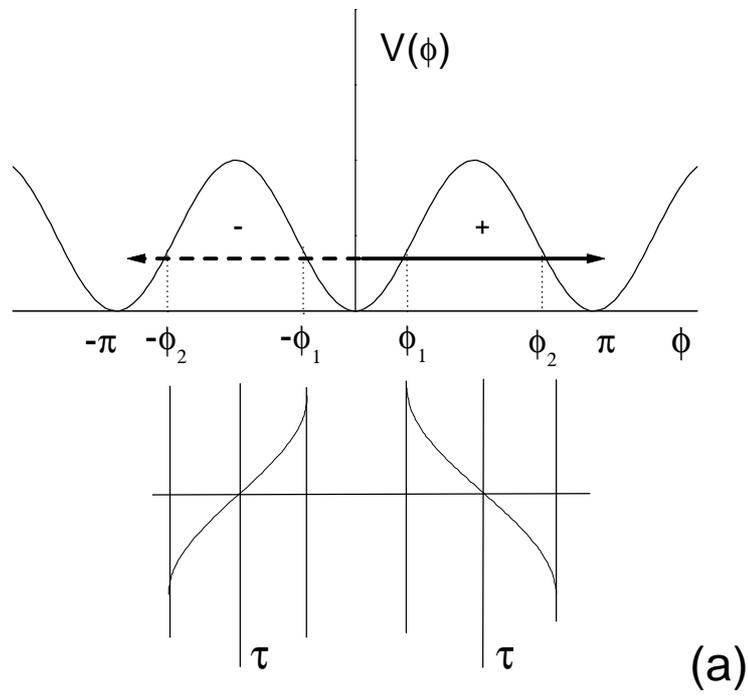
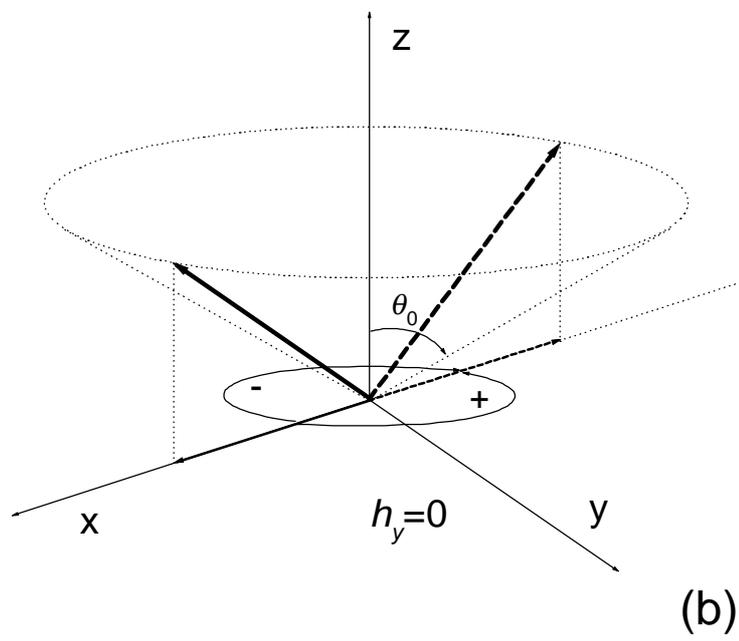
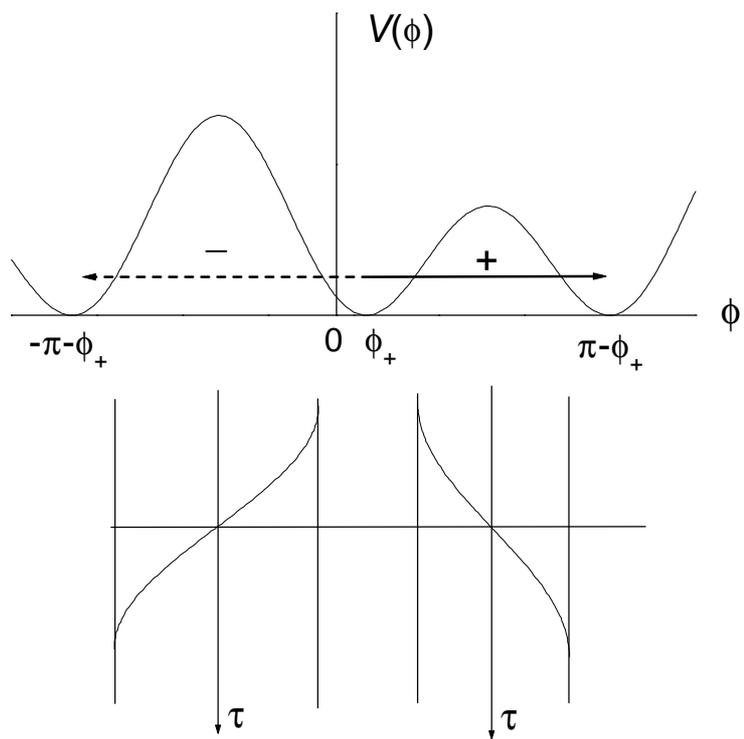
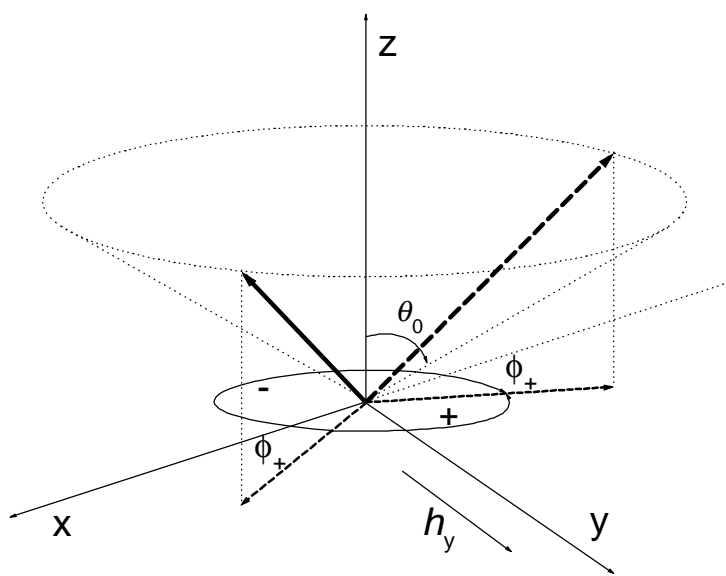
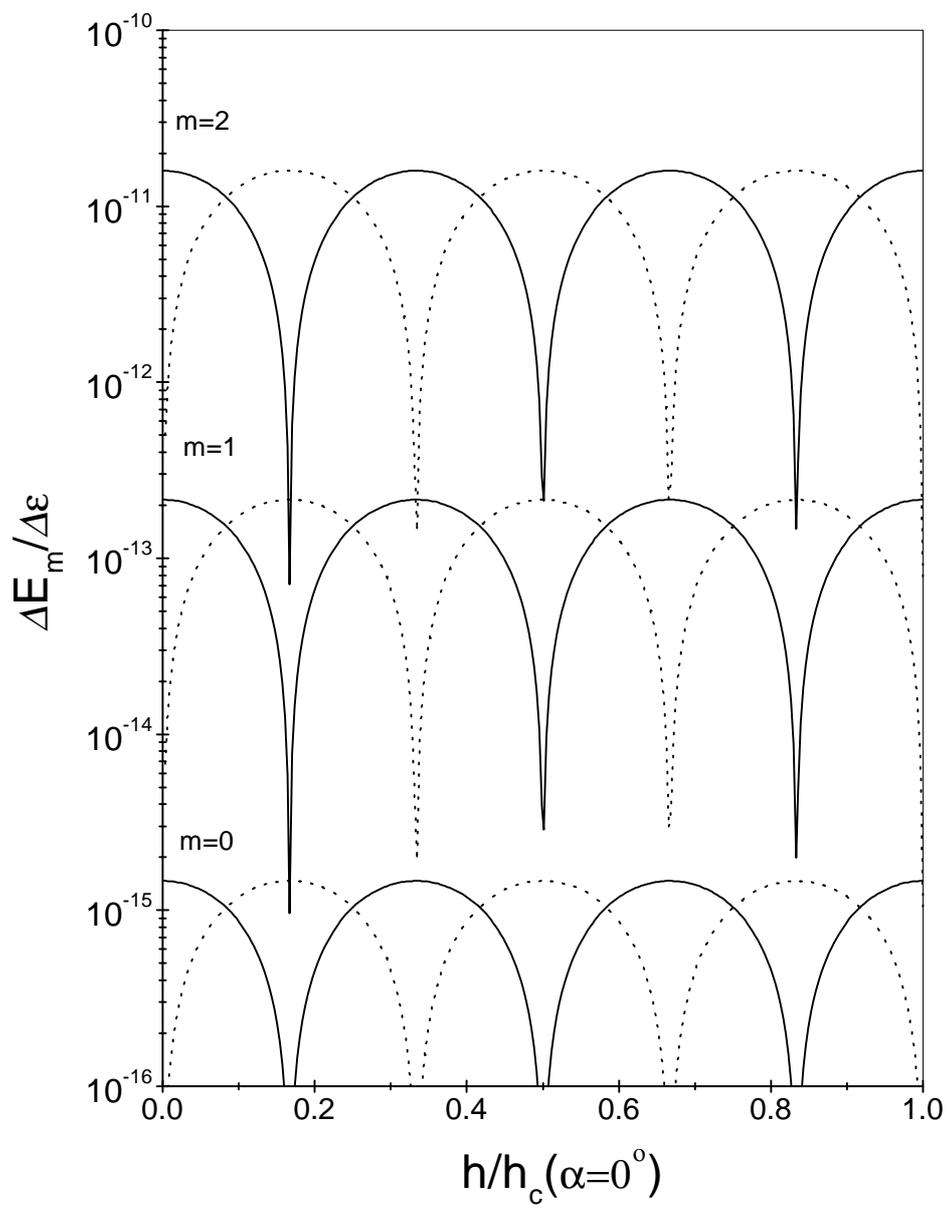
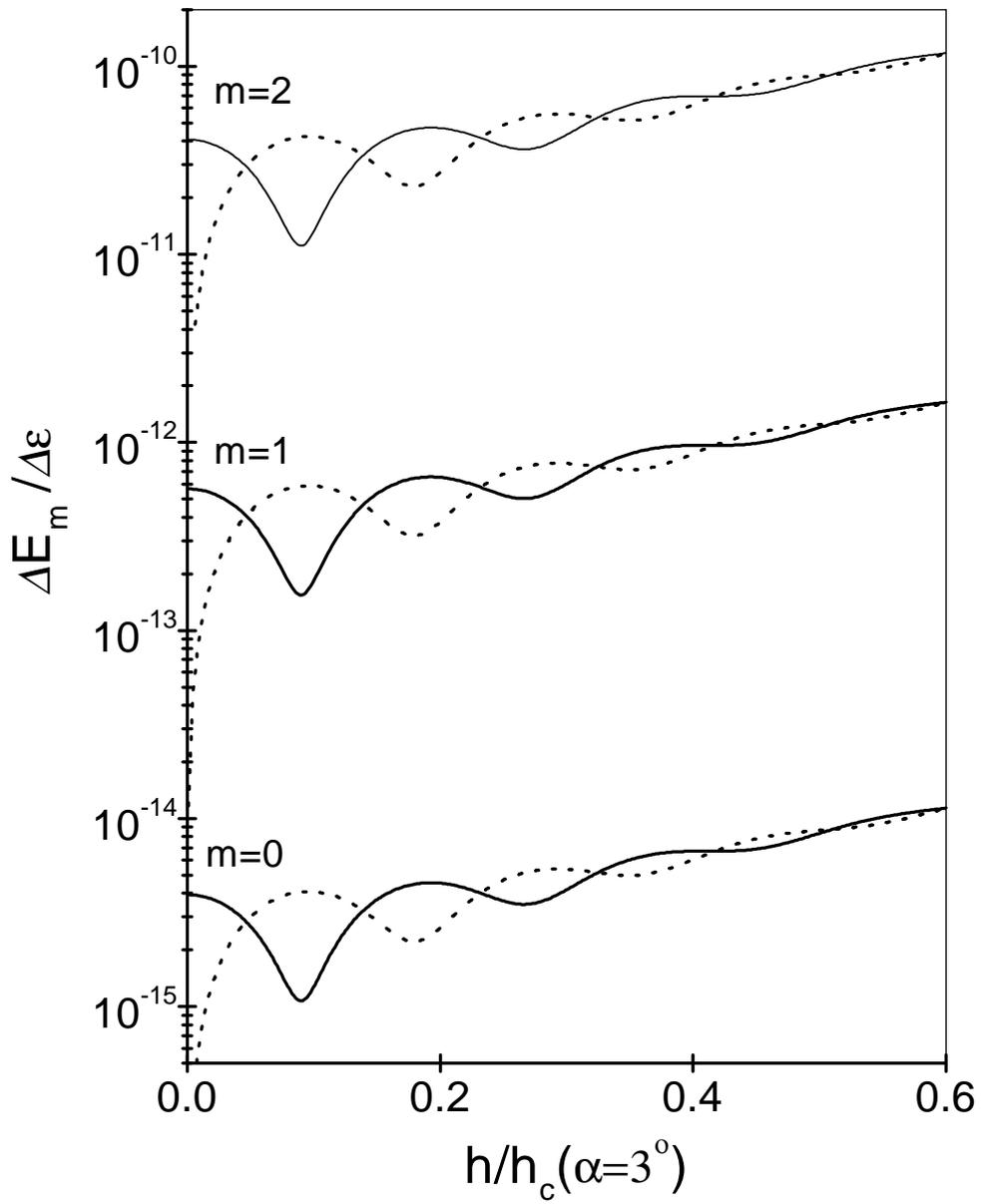
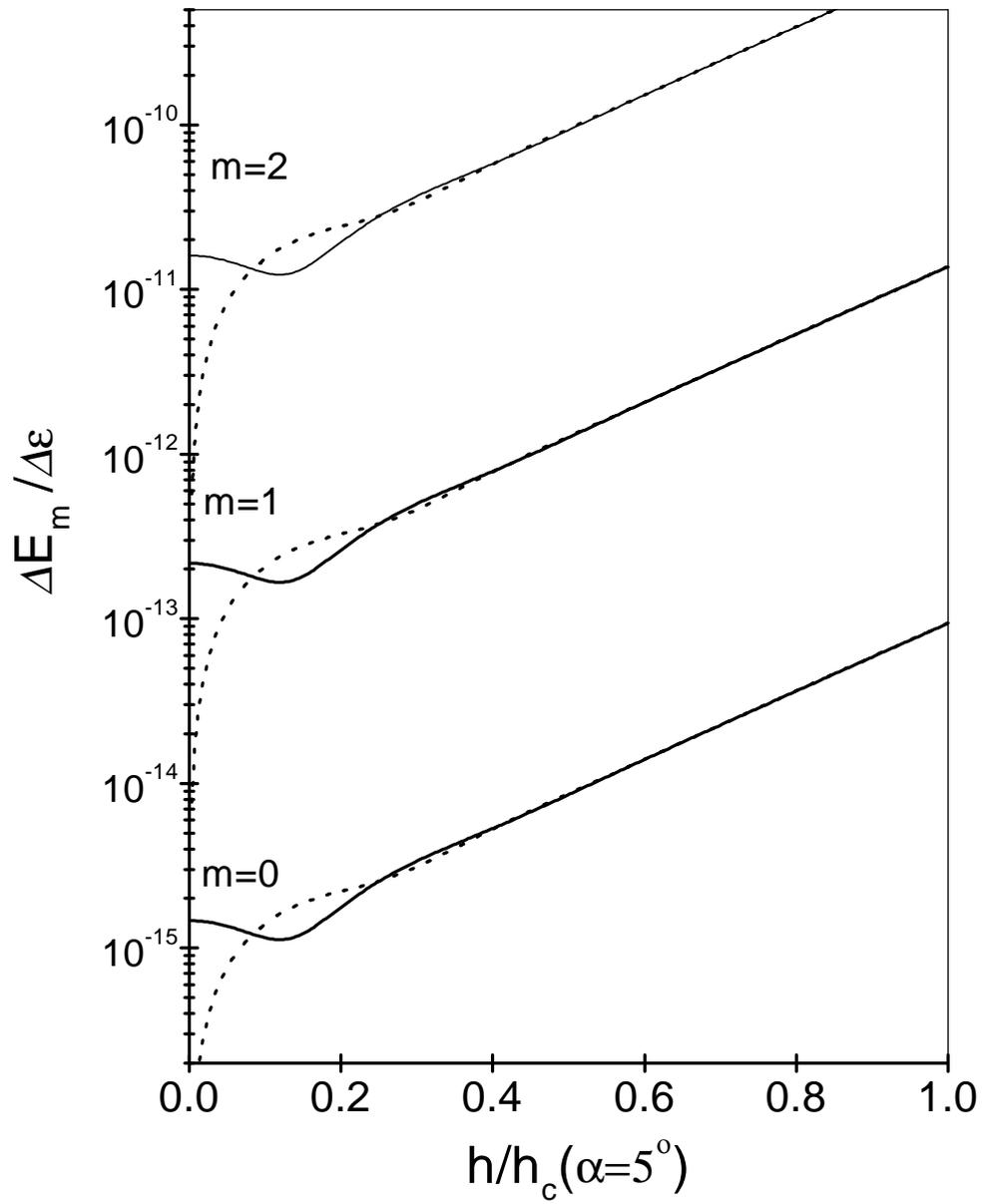
\end{document}